\documentclass[%
 reprint,
superscriptaddress,
 amsmath,amssymb,
 aps,
 pra,
 longbibliography
]{revtex4-2}
\pdfoutput=1
\usepackage[utf8]{inputenc}
\usepackage[english]{babel}
\usepackage[T1]{fontenc}
\usepackage{amsmath}
\usepackage{hyperref}
\usepackage{mathrsfs}

\usepackage{url}

\usepackage{pgfplots, pgfplotstable}

\usepackage{tikz}
\usepackage{physics}
\usepackage{amsfonts}
\usepackage[utf8]{inputenc}
\usepackage{grffile}
\usepackage{graphicx}
\usepackage{flafter}
\usepackage{siunitx}
\usepackage[nameinlink]{cleveref}
\usepackage{dcolumn}
\usepackage{bm}
\usepackage{tikz}
\usetikzlibrary{quantikz}
\usetikzlibrary{shapes.geometric}
\tikzstyle{tensor}=[minimum size=4mm,draw=blue!100,fill=blue!20,thick]
\tikzstyle{tensor2}=[circle,minimum size=4mm,draw=gray!100,fill=gray!20,thick]
\tikzstyle{tensor-expanded}=[rounded corners,minimum width=14mm, minimum height=4mm,draw=gray!100,fill=gray!70,thick]

\tikzstyle{RBM-visible}=[circle,minimum size=4mm,draw=gray!100,fill=gray!20,thick]

\tikzstyle{RBM-hidden}=[circle,minimum size=4mm,draw=blue!100,fill=blue!20,thick]
\usepackage[caption=false]{subfig}

\usepackage{xcolor}
\usepackage[normalem]{ulem}
\definecolor{mygreen}{rgb}{0.21, 0.73, 0.13}

\begin{document}
\title{Quantum resources of
quantum and classical variational methods}

\begin{abstract}
Variational techniques have long been at the heart of atomic, solid-state, and many-body physics. They have recently extended to quantum and classical machine learning, providing a basis for representing quantum states via neural networks. These methods generally aim to minimize the energy of a given ansatz, though open questions remain about the expressivity of quantum and classical variational ansätze. The connection between variational techniques and quantum computing, through variational quantum algorithms, offers opportunities to explore the quantum complexity of classical methods. We demonstrate how the concept of non-stabilizerness, or magic, can create a bridge between quantum information and variational techniques and we show that energy accuracy is a necessary but not always sufficient condition for accuracy in non-stabilizerness. Through systematic benchmarking of neural network quantum states, matrix product states, and variational quantum methods, we show that while classical techniques are more accurate in non-stabilizerness, not accounting for the symmetries of the system can have a severe impact on this accuracy. Our findings form a basis for a universal expressivity characterization of both quantum and classical variational methods.
\end{abstract}

\author{Thomas Spriggs}
\thanks{T.S. and A.A. contributed equally}
\email{t.e.spriggs@tudelft.nl and a.ahmadi-1@tudelft.nl}
\affiliation{QuTech and Kavli Institute of Nanoscience, Delft University of Technology, Delft, the Netherlands}

\author{Arash Ahmadi}
\thanks{T.S. and A.A. contributed equally}
\email{t.e.spriggs@tudelft.nl and a.ahmadi-1@tudelft.nl}
\affiliation{QuTech and Kavli Institute of Nanoscience, Delft University of Technology, Delft, the Netherlands}

\author{Bokai Chen}
\affiliation{Kavli Institute of Nanoscience, Delft University of Technology, Delft, the Netherlands}

\author{Eliska Greplova}
\affiliation{QuTech and Kavli Institute of Nanoscience, Delft University of Technology, Delft, the Netherlands}

\maketitle

\section{Introduction}

Due to the prohibitive scaling of large scale quantum wavefunctions, approximate methods are typically employed to numerically find ground states and time evolve quantum systems. Classical variational methods have rich track record and span physics intuition driven parametrization of the trial wavefunction \cite{Jastrow_1955,PhysRevLett.10.159,PhysRev.134.A923,PhysRev.137.A1726,PhysRev.108.1175}, ansätze capturing specific entanglement properties \cite{SCHOLLWOCK201196,perez2007peps}, and, more recently, physics-agnostic wavefunctions based on neural networks \cite{carleo_RBM_2017,PhysRevB.100.125124, Hibat-Allah_Carrasquilla_2020, Viteritti:2022fji, Luo:2020stn, roth2021group}.

In parallel to this progress, quantum variational methods were formulated with the same goal: to capture potentially complicated wavefunctions using a small number of degrees of freedom~\cite{cerezo2021variational,kremenetski2108quantum,moll2018quantum,yao2021adaptive}. In this instance, the variational ansatz is formulated as a quantum circuit. Quantum operations in such circuits are then parametrized and the parameters of the quantum gates are updated, using classical optimization techniques, until the energy of the output of the circuit well approximates the energy of the sought-after quantum state. 

In search for ground states of specific Hamiltonians, both classical and quantum variational techniques rely on the energy variational principle: the expectation value of the Hamiltonian is evaluated with respect to the trial state and then minimized. The lowest achievable value is then considered a good approximation for the true ground state energy of the system.

Once we optimize the energy of our ansatz, the question arises regarding how well is the quantum state itself represented by our approximate representation. While variational minimization brings the system to the point in Hilbert space with the required energy, it is far from certain that the optimized state will structurally reflect the global properties of the system. In particular, especially long range correlations are known to pose a challenge.

In this work, we assess the expressivity of both quantum and classical variational methods from a quantum computational complexity point of view: we analyze the amount of quantum resources that the variational ansatz expresses when optimized in accordance with the classical energy variational principle. 

In quantum computing, one way of assessing quantum resources is to measure how far a given state is from being efficiently and exactly simulated on a classical computer \cite{gottesman1998}. The quantum operations that allow for efficient, exact classical simulation are elements of the Clifford group. All the other operations are referred to as non-Clifford. To quantify how `far' a given operation is from the Clifford group, a measure called magic, or non-stabilizerness, was introduced~\cite{Bravyi_Kitaev2005}.

Non-stabilizerness has recently been at the forefront of quantum information literature as more scalable techniques have emerged to quantitatively evaluate it \cite{haug2023efficient,haug_mps_2023,tarabunga2024nonstabilizerness,lami_mps_2023,ahmadi2024quantifying}. 
This progress allowed for the first exploration of non-stabilizerness evaluation for tensor networks~\cite{tarabunga2024nonstabilizerness,frau2024nonstabilizerness,White_2021}.

At first sight, the notions of non-stabilizerness and classical variational principle are unrelated. One is designed to determine areas of potential quantum-computational advantage, the other to asses performance of classical, few-parameter ansätze. However, a key property we want from a variational ansatz is to faithfully approximate the sought-after state, beyond just its energy. An interesting interplay with quantum information arises: classical variational ansätze are unrestricted by, for example, the Gottesmann-Knill theorem~\cite{gottesman1998}, and are allowed to converge to any point in the Hilbert space that minimize the energy regardless of where the exact solution lies with respect to the Clifford group.

In this work, we perform a systematic benchmark for non-stabilizerness expressivity of classical and quantum variational ansätze using the transverse-field Ising model (TFIM), a common testbed model for variational methods. Specifically, we compare non-stabilizerness expressivity of neural quantum states, matrix product states, and a variational quantum eigensolver.
We find that high energy precision is a necessary, but not always sufficient, condition for the few-parameter variational ansätze to express the non-stabilizerness of the model correctly.
We observe that quantum variational methods generally show worse performance compared to the state-of-the-art classical methods. In the context of this work it means that even though the ansatz itself is quantum, it is worse at predicting the energy and non-stabilizerness of the exact wavefunction. Moreover, the variations in energy and non-stabilizerness accuracy across multiple repeats are generally much larger for the quantum variational method compared to the results from neural quantum states.

The outline of this paper is as follows. Section~\ref{sec:methods} provides an overview of methods and techniques used in this work: Section~\ref{sec:TFIM} briefly elucidates the specific Hamiltonian with which this work explores, Section \ref{sec:param_QS} details the three algorithms used for finding the ground state of the given Hamiltonian, and Section \ref{sec:non-stab} presents the measure of quantum resources that we use to assess the non-stabilizerness of the ground state wavefunctions obtained. Section \ref{sec:results} presents the comparison of the ground state energy and non-stabilizerness found using density matrix renormalization group (DMRG), neural quantum states (NQS), and a variational quantum eigensolver (VQE); this comparison focuses on the respective accuracies of different methods compared to exact diagonalization, the interplay between the energy and non-stabilizerness accuracies, and the fluctuations of their solutions with repeated runs. In Section~\ref{sec:conclusions} we discuss our findings and possible next steps in aligning classical and quantum variational methods.

\section{Methods}
\label{sec:methods}
\subsection{Transverse-Field Ising Model} \label{sec:TFIM}

This paper is focused on parameterized ansätze that approximate the ground state of the transverse-field Ising model (TFIM).  The Hamiltonian of this model is defined as
\begin{equation}
    H = J\sum \sigma^z_i \sigma^z_{i+1} - h\sum \sigma^x_i. 
    \label{eq:TFIM}
\end{equation}
In this definition, $\sigma^z$ and $\sigma^x$ are the Pauli matrices, $J$ is a constant that dictates the coupling between neighboring spins (we will consider $J<0$ and thus neighboring spins being aligned is energetically favorable), and $h$ is the transverse magnetic field strength; we are considering the case of periodic boundaries. The basis that will be used to represent the wavefunction will be the spin projections along the $z$-axis, comprised of either spin up or down relative to this axis. This is also known as the computational basis.

The TFIM begets a phase transition with a critical point in the vicinity of  $|h/J| = 1$~\cite{sachdev2023quantum}. To explore this phase diagram we will find solutions of the ground state with fixed $J=-1$ and varying $h$ over the range of $0 - 3$.

\subsection{Parameterized Quantum States} \label{sec:param_QS}

\begin{figure}
    \centering
    \begin{tikzpicture}[inner sep=1mm]
    \tikzstyle{every line}=[line width=0.5pt]

            \node[RBM-visible] (v1) at (1, 1.75) {$\sigma^z_1$};
            \node[RBM-visible] (v2) at (2.5, 1.75) {$\sigma^z_2$};
            \node[RBM-visible] (v3) at (4, 1.75) {$\sigma^z_3$};
            \node[RBM-visible] (vN) at (6, 1.75) {$\sigma^z_N$};
            \node[RBM-hidden] (h1) at (0, 0) {$h_1$};
            \node[RBM-hidden] (h2) at (1.5, 0) {$h_2$};
            \node[RBM-hidden] (h3) at (3, 0) {$h_3$};
            \node[RBM-hidden] (hM) at (7, 0) {$h_M$};

            \draw[every line] (v1) -- (h1);         
            \draw[every line] (v1) -- (h2);         
            \draw[every line] (v1) -- (h3);         
            \draw[every line] (v1) -- (hM);    
            
            \draw[every line] (v2) -- (h1);         
            \draw[every line] (v2) -- (h2);         
            \draw[every line] (v2) -- (h3);         
            \draw[every line] (v2) -- (hM);      
            
            \draw[every line] (v3) -- (h1);         
            \draw[every line] (v3) -- (h2);         
            \draw[every line] (v3) -- (h3);         
            \draw[every line] (v3) -- (hM);   
            
            \draw[every line] (vN) -- (h1);         
            \draw[every line] (vN) -- (h2);         
            \draw[every line] (vN) -- (h3);         
            \draw[every line] (vN) -- (hM);        

            \node [at={(0.12,1.25)}, align=center] {$W_{11}$}; 
            
            \node [at={(5,1.75)}, align=center] {...};    
            \node [at={(5,0)}, align=center] {...};    
            
            \end{tikzpicture}   
    \caption{A pictorial representation of the restricted Boltzmann machine (RBM) with $M$ hidden neurons and $N$ visible neurons encoding the projection in the $\sigma^z$ basis.}
    \label{fig:RBM_rep}
\end{figure}

The energy spectrum, dynamics, and a plethora of other observables related to a quantum system are obtainable through the wavefunction, $\psi$. However, Nature does not grant one access to the wavefunction, nor would it be computationally feasible to store these data for a system larger than a few tens of qubits. The latter complication is due to the exponential increase in the number eigenstates with system size. In this work we employ three different parameterized quantum states that all circumvent these limitations.

Parameterized quantum states can offer two key features. Firstly, by representing the wavefunction as a parameterized function, $\psi_{\boldsymbol{\theta}}$, we can learn a representation of the true wavefunction by learning the parameters $\boldsymbol{\theta}$. Secondly, by restricting the representation such that the number of parameters grows polynomially (rather than exponentially) with the system size, we achieve an efficient representation that makes simulations feasible for larger systems.

\subsubsection{Neural Quantum States} \label{sec:nqs}
Beginning with the work of Carleo and Troyer, neural-network-based methods were introduced to act as physics-agnostic, parameterized quantum states known as neural quantum states (NQS) \cite{carleo_RBM_2017}. NQS have shown to be useful in a variety of problems, such as finding the ground state in different quantum systems \cite{Nomura_Imada_2021,roth2021group,Astrakhantsev_Westerhout_2021}, evolving quantum states through time \cite{carleo_RBM_2017,Gutierrez2022realtimeevolution}, and simulating quantum circuits \cite{Medvidovi__2021}. There exist different architectures for the NQS \cite{carleo_RBM_2017,PhysRevB.100.125124,Hibat-Allah_Carrasquilla_2020,Luo:2020stn,roth2021group,Viteritti:2022fji}, for the purpose of this paper, however, we restrict ourselves to a simple case of NQS, namely the restricted Boltzmann machine (RBM) depicted in Figure \ref{fig:RBM_rep}. 

The RBM has the simple structure of two neural network layers with no intra-layer connections but full inter-layer connectivity: one visible layer consisting of $N$ units and one hidden layer of $M$ units. $N$ must be equal to the number of qubits/particles in the system, whereas there is no formal restriction on $M$; the ratio $M/N$, referred to as $\alpha$, is often used to convey the size of an RBM. The number of parameters in the RBM is given by $N + \alpha N + \alpha N^2$, and thus for fixed $\alpha$ this scales polynomially with the system size; however, this does not guarantee that the desired state can be represented by this network, in which case the true ground state may still require a number of parameters larger than this. The wavefunction amplitude given by the RBM ansatz is 
\begin{equation}
    \psi_{\boldsymbol{\theta}} (\mathbf{s}) = \sum_\mathbf{h} e^{\sum_j a_j \sigma_j +\sum_i b_i h_i+\sum_{ij}W_{ij}h_i \sigma_j},
    \label{eq:RBM_amplitude}
\end{equation}
where $\mathbf{s}=(\sigma_1,\sigma_2,...,\sigma_N)$ such that $\sigma_i= \pm 1$ is the spin configuration in $z$ basis; $\mathbf{h} = (h_1,h_2,...,h_M)$, with $h_i=\pm 1$, denotes the $M$ hidden spin variables; and the parameters to be optimized are $\boldsymbol{\theta} = \{\boldsymbol{a},\boldsymbol{b},\boldsymbol{W}\}$. The full wavefunction can be constructed from 
\begin{equation}
    \ket{\psi_{\boldsymbol{\theta}}} = \sum_\textbf{s}\psi_{\boldsymbol{\theta}} (\mathbf{s})\ket{\textbf{s}}.
\end{equation} 

With the specific form of the parameterized quantum state in-hand, what remains is to train this model such that the wavefunction that is represented captures the behavior of the ground state of the Hamiltonian being considered. To do this we will consider the expectation value of the Hamiltonian with respect to the NQS wavefunction, $\langle H \rangle_{\psi_{\boldsymbol{\theta}}}:= \bra{\psi_{\boldsymbol{\theta}}} H \ket{\psi_{\boldsymbol{\theta}}}$. In line with the variational principle~\cite{becca_sorella_2017}, this quantity is bounded from below by the ground state energy and will equal that bound when the NQS represents the ground state wavefunction.

One of the benefits of a parameterized quantum state is that, for a fixed $\alpha$, the number of free parameters that characterize the wavefunction in Equation \eqref{eq:RBM_amplitude} is polynomial with the number of qubits. This work would be undone if we required the full wavefunction or access to the entire Hilbert space to compute expectation values.

To maintain the efficient, computationally feasible computations of expectation values we use Monte Carlo (MC) sampling alongside so-called \textit{local estimators}. The combination of MC sampling and a variational ansatz is known as variational Monte Carlo (VMC). An excellent outline of the computational efficiencies included in VMC is outlined in \cite{Medvidovic:2024ihh}. Following the conventions of that work, we can reduce the calculation of the expectation value from a sum over an exponentially large Hilbert space to a statistical expectation of local estimators. Thus reducing 
\begin{equation}
    \langle \hat{O} \rangle = \frac{\bra{\psi_{\boldsymbol{\theta}}} \hat{O} \ket{\psi_{\boldsymbol{\theta}}}}{\braket{\psi_{\boldsymbol{\theta}}}},
    \label{eq:expectation}
\end{equation}
to the more tractable sum of $\langle \hat{O} \rangle = \langle O_{\text{loc}} \rangle_{P}$. The local estimator and probability distribution are defined as 
\begin{equation}
    O_{\text{loc}}(\textbf{s}) = \sum_{\textbf{s}'} \bra{\textbf{s}} \hat{O} \ket{\textbf{s}'} \frac{\bra{\textbf{s}'}\ket{\psi_{\boldsymbol{\theta}}}}{\bra{\textbf{s}}\ket{\psi_{\boldsymbol{\theta}}}},
    \label{eq:local-operator}
\end{equation}
and
\begin{equation}
    P(\textbf{s}) = \frac{|\bra{\psi_{\boldsymbol{\theta}}} \ket{\textbf{s}}|^2}{\sum_{\textbf{s}'} |\bra{\psi_{\boldsymbol{\theta}}} \ket{\textbf{s}'}|^2},
    \label{eq:MC_sampling_P}
\end{equation}
respectively. The VMC procedure generates an ensemble of configurations distributed under $P(\textbf{s})$. 

The computation of Equation \eqref{eq:local-operator} is only efficient if each local estimator contains, at most, a polynomial number (relative to the total number of qubits in the system) of connected states, $\textbf{s}'$: equivalent to only having a polynomial number of distinct matrix elements, $ \bra{\textbf{s}} \hat{O} \ket{\textbf{s}'}$. As can be seen from Equation \eqref{eq:TFIM}, the Hamiltonian of the TFIM contains only terms that involve one or two qubits, and thus there are only two or four connected states for each term in the Hamiltonian, regardless of the system size. Therefore, the condition of only a polynomial number of neighboring states is met. 

Using this efficient computation for expectation values, we can compute the energy of the NQS for a given set of parameters, $\langle H \rangle_{\psi_{\boldsymbol{\theta}}}$. As the NQS is a differentiable function, we can also compute the gradient of this energy with respect to each of the network's parameters. As is the common paradigm for training neural networks, the cost function (in this case the energy) and its derivatives are used to tune the parameters until the cost function is minimized. And, as previously mentioned, this will occur when the NQS represents the ground state wavefunction. For this work we used the Python package \texttt{NetKet} \cite{netket2:2019,netket3:2022} which implements the efficient computations laid out here.

For small systems, the entire Hilbert space is not prohibitively large. Therefore, for comparison, we can compute the exact expectation value, $\langle H \rangle_{\psi_{\boldsymbol{\theta}}}$, rather than having to rely on MC sampling. The effects of exact against MC sampling is mentioned in Appendix \ref{sec:RBM_appendix} alongside further details of the RBM architecture. Unless stated otherwise, all RBM data shown in this paper was generated from a network with $\alpha=5$.

\subsubsection{Density Matrix Renormalization Group} \label{sec:DMRG}
A well established, classical approach to finding the smallest eigenvalue of a given Hamiltonian was introduced by White in \cite{White:1992zz}, and is known as DMRG. DMRG is an iterative process of solving smaller matrix inversion problems to eventually invert a larger matrix. If the inversion problem can be written in terms of matrix product states (MPS) and matrix product operators (MPO) then there is an efficient implementation of DMRG \cite{SCHOLLWOCK201196}. 

A generic tensor can be represented as an MPS or MPO. However, tensors representing low-entanglement wavefunctions, or local Hamiltonians, can be represented as low-rank MPSs and MPOs respectively. The rank of the MPS is called the bond dimension, $D$, and can be adjusted during the DMRG procedure to balance between computational efficiency (small $D$) and allowing for high entanglement (large $D$). It is common practice to limit the bond dimension to achieve a more efficient representation of the wavefunction \cite{Cirac_2021}. For this work, however, we allowed the DMRG algorithm to grow the bond dimension to as large a value as 100 if there was a significant decrease in energy.

\begin{figure}
    \begin{tikzpicture}[inner sep=1mm]
    \node (Hpsi) at (0,2.7) {
    \begin{tikzpicture}[inner sep=1mm]
    \tikzstyle{every line}=[line width=2pt]

            \node[tensor] (0) at (0, 0) { };
            \node[tensor] (1) at (1, 0) { };
            \node[tensor] (2) at (2, 0) { };
            \node[tensor] (3) at (3, 0) { };
            
            \node (0spin) at (0, -0.5) { };
            \node (1spin) at (1, -0.5) { };
            \node (2spin) at (2, -0.5) { };
            \node (3spin) at (3, -0.5) { };

            \node[tensor2] (0MPO) at (0, +1) { };
            \node[tensor2] (1MPO) at (1, +1) { };
            \node[tensor2] (2MPO) at (2, +1) { };
            \node[tensor2] (3MPO) at (3, +1) { };

            \draw[every line] (0) -- (0MPO);
            \draw[every line] (1) -- (1MPO);
            \draw[every line] (2) -- (2MPO);
            \draw[every line] (3) -- (3MPO);
            
            \draw[every line] (0) -- (0spin);            
            \draw[every line] (1) -- (1spin);            
            \draw[every line] (2) -- (2spin);            
            \draw[every line] (3) -- (3spin);            

            \draw[every line] (0) -- (1);            
            \draw[every line] (1) -- (2);            
            \draw[every line] (2) -- (3);            

            \draw[every line] (0MPO) -- (1MPO);            
            \draw[every line] (1MPO) -- (2MPO);            
            \draw[every line] (2MPO) -- (3MPO);            
            \end{tikzpicture}
    };

    \node (Epsi) at (4.5,2.7) {
    \begin{tikzpicture}[inner sep=1mm]
    \tikzstyle{every line}=[line width=2pt]

            \node[tensor2] (0) at (0, 0) { };
            \node[tensor2] (1) at (1, 0) { };
            \node[tensor2] (2) at (2, 0) { };
            \node[tensor2] (3) at (3, 0) { };
            
            \node (0spin) at (0, -0.5) { };
            \node (1spin) at (1, -0.5) { };
            \node (2spin) at (2, -0.5) { };
            \node (3spin) at (3, -0.5) { };

            \draw[every line] (0) -- (0spin);            
            \draw[every line] (1) -- (1spin);            
            \draw[every line] (2) -- (2spin);            
            \draw[every line] (3) -- (3spin);            

            \draw[every line] (0) -- (1);            
            \draw[every line] (1) -- (2);            
            \draw[every line] (2) -- (3);            

            \end{tikzpicture}
    };

    \node (psiHpsi) at (0.5,0) {
    \begin{tikzpicture}[inner sep=1mm]
    \tikzstyle{every line}=[line width=2pt]

            \node[tensor] (0MPO) at (0, 0) { };
            \node[tensor] (1MPO) at (1, 0) { };
            \node[tensor] (2MPO) at (2, 0) { };
            \node[tensor] (3MPO) at (3, 0) { };

            \node[tensor2] (0MPS) at (0, +1) { };
            \coordinate (1MPS) at (1, +1) { };
            \coordinate (left1MPS) at (0.5,1);
            \coordinate (under1MPS) at (1,0.5);
            \coordinate (right2MPS) at (2.5,1);
            \coordinate (under2MPS) at (2,0.5);

            \node[tensor2] (3MPS) at (3, +1) { };

            \node[tensor2] (0MPS2) at (0, -1) { };
            \node[tensor2] (1MPS2) at (1, -1) { };
            \node[tensor2] (2MPS2) at (2, -1) { };
            \node[tensor-expanded] (2MPS2-long) at (1.5, -1) { };
            \node[tensor2] (3MPS2) at (3, -1) { };

            \draw[every line] (0MPS) -- (0MPO);
            \draw[every line] (under1MPS) -- (1MPO);
            \draw[every line] (under2MPS) -- (2MPO);
            \draw[every line] (3MPS) -- (3MPO);
         
            \draw[every line] (0MPO) -- (0MPS2);
            \draw[every line] (1MPO) -- (1MPS2);
            \draw[every line] (2MPO) -- (2MPS2);
            \draw[every line] (3MPO) -- (3MPS2);

            \draw[every line] (0MPS) -- (left1MPS);            
            \draw[every line] (right2MPS) -- (3MPS);     
            
            \draw[every line] (0MPS2) -- (1MPS2);            
            \draw[every line] (2MPS2) -- (3MPS2);            

            \draw[every line] (0MPO) -- (1MPO);            
            \draw[every line] (1MPO) -- (2MPO);            
            \draw[every line] (2MPO) -- (3MPO);            
            \end{tikzpicture}
    };

        \node (psiHpsi-expanded-node) at (4.1,0) {
    \begin{tikzpicture}[inner sep=1mm]
    \tikzstyle{every line}=[line width=2pt]

            \coordinate (leftMPS) at (0,1);
            \coordinate (rightMPS) at (2,1);
            \node[tensor2] (halfrightMPS) at (1.5,1) { };
            \node[tensor2] (halfleftMPS) at (0.5,1) { };
            \coordinate (aboveleftMPS) at (0.5,1.5);
            \coordinate (aboverightMPS) at (1.5,1.5);
            \node[tensor-expanded] (MPS) at (1, +1) { };

            \draw[every line] (MPS) -- (leftMPS);
            \draw[every line] (halfrightMPS) -- (aboverightMPS);
            \draw[every line] (halfleftMPS) -- (aboveleftMPS);
            \draw[every line] (MPS) -- (rightMPS);
        
            \end{tikzpicture}
    };

    \draw (Hpsi) (Epsi);
    \node at (2.2,2.9) [scale=1.3] {= $E_0$};
    \node at (-2,4) [scale=1] {(a)};

    \draw (Hpsi) (psiHpsi-expanded-node);
    \node at (-2,1.4) [scale=1] {(b)};
    \node at (2.7,-0.1) [scale=1.3] {- $\lambda$};
    \node at (5.6,-0.1) [scale=1.3] {= 0};

    \end{tikzpicture}
    \caption{Tensor network representation of the eigenvalue equation $H\ket{\psi} = E_0\ket{\psi}$ shown in (a). The gray nodes represent the MPS version of the ground state wavefunction, and the blue nodes are the Hamiltonian converted into an MPO. Panel (b) shows an intermediate stage of the DMRG algorithm which poses a smaller inversion problem whereby a pair of nodes are discarded and only the dark gray node is optimized to minimize $\lambda$; the dark node is created by contracting the two neighbouring nodes at the sites straddled by this dark gray node. The location of the node that is updated sweeps left and right along the chain until some convergence criterion is met.}
    \label{fig:DMRG-MPS}
\end{figure}

The parameters that get updated in the DMRG algorithm are the elements of the MPS at each node. The way that these parameters are optimized is as follows:
\begin{enumerate}
    \item An initial, random, set of values is chosen for every element in the MPS $\ket{\psi}$.
    \item The MPO representing the Hamiltonian is contracted from both sides by the MPS. This is equivalent to the equation $\bra{\psi}H\ket{\psi} = E_0$.
    \item A pair of neighboring nodes are chosen and are contracted into a larger tensor; the result is no longer an MPS (this is shown in Figure \ref{fig:DMRG-MPS}(b)). An iterative method is then performed in this subspace (subspace because it is still not the entire Hilbert space) to compute a new tensor that replaces the selected pair but lowering the eigenvalue of the MPS.
    \item The altered tensor is then broken back into two tensors using singular value decomposition (SVD) to return $\ket{\psi}$ to an MPS whilst maintaining the lower eigenvalue achieved in step 3.
    \item The next neighboring pair of links are chosen and steps 3 and 4 are repeated, systematically moving along the chain of the MPS, until a predefined convergence criterion is met.
\end{enumerate}

Upon the termination of the DMRG algorithm, one is left with an MPS that represents the ground state wavefunction, and the ground state energy. For this work we used the implementation provided by \texttt{ITensor} \cite{10.21468/SciPostPhysCodeb.4,10.21468/SciPostPhysCodeb.4-r0.3}.

\subsubsection{Variational Quantum Eigensolver}
An alternate approach to classically approximating the wavefunction is to use a parameterized quantum circuit to prepare a trial wavefunction and then use classical optimization to vary the parameters until a desired wavefunction is reproduced. When this process is applied to finding the ground state wavefunction it is known as a variational quantum eigensolver (VQE) \cite{mcclean2016theory}. 

The formulation of VQE spearheaded the development of a rich field of variational quantum algorithms (VQAs)~\cite{cerezo2021variational,kremenetski2108quantum,moll2018quantum,yao2021adaptive}. We will first outline the VQE algorithm and then explore some of the challenges related to its implementation.

One begins the VQE procedure with an easily prepared initial state wavefunction, $\ket{\psi_{in}}$, to which one applies a sequence of unitary operators, or \textit{gates}, that comprise the parameterized quantum circuit. The output of the circuit will be a parameterized wavefunction $\ket{\psi_{\boldsymbol{\theta}}}$. Thus, with the quantum circuit denoted by the unitary matrix $\mathcal{U}(\boldsymbol{\theta})$, the parameterized wavefunction generated by the quantum circuit reads
\begin{equation}
    \ket{\psi_{\boldsymbol{\theta}}} = \mathcal{U}(\boldsymbol{\theta})\ket{\psi_{in}}.
\end{equation}
An example of a parameterized quantum circuit is depicted in Figure \ref{fig:VQE_circuit_1_layer}; the circuit used for this work is four contiguous layers of the circuit shown, repeated in series. The values of the parameterized gates are varied to find the minimum of the expectation value of the Hamiltonian, Equation \eqref{eq:TFIM}, with respect to  $\ket{\psi_{\boldsymbol{\theta}}}$, denoted $\langle H \rangle_{\psi_{\boldsymbol{\theta}}}$ as in Section \ref{sec:nqs}. For this work, given the system sizes simulated, a classical simulation of the quantum circuit was performed using \texttt{Pennylane} \cite{Bergholm:2018cyq}. There are, however, implementations on quantum hardware \cite{peruzzo2014variational,o2016scalable,kandala2017hardware,10313698,Atas:2021ext,Zhang:2021bjq}. 

Given that this work presents classical simulations of quantum circuits, access is granted to the full wavefunction rather than just projective measurements. This limits the size of the system that can be simulated but the effects of using the full wavefunction rather than projective measurements to estimate $\langle H \rangle_{\psi_{\boldsymbol{\theta}}}$ is presented in Appendix \ref{sec:VQE_appendix}. 

Considering, still, the case of a classical simulation, free from hardware-related noise and imperfections, there exist fundamental limitations in the VQE algorithm. Firstly, common to all variational algorithms that require a predetermined ansatz, including those in this Section, is that there is no guarantee that the ansatz in question (the quantum circuit) is expressive enough to represent the function it hopes to capture (the wavefunction) using a polynomial number of parameters. This is especially true for strongly correlated quantum systems~\cite{avella2012strongly}. Broadly speaking there are two approaches to circuit design for VQEs: problem-specific ansätze and hardware-efficient ansätze (HEA). In the former regime, one constructs the ansatz from elements that ones hopes captures known behaviors of the solution. The latter attempts to create a more general, easier to implement on quantum hardware, ansatz built from simple gates; this is where the ansatz for this work resides. Neither approach, however, is guaranteed to produce and ansatz that can represent the desired function, and thus the algorithm can converge on a suboptimal solution. Secondly, VQAs have been shown to be especially prone to so-called \textit{barren plateaus}~\cite{mcclean2018barren,wang2021noise,cerezo2022challenges,PRXQuantum.3.010313}. The phenomenon of a barren plateau is when the gradient of the loss function with respect to the variational parameters vanishes exponentially with the number of qubits in the system. This means that training the variational parameters becomes exceptionally difficult for even modest sized systems - even with gradient-free optimizers \cite{arrasmith2021effect} - as the true minimum becomes exponentially hard to find. Thirdly, and even affecting small systems, the loss landscape for VQEs often contains many local minima, and therefore even without vanishing gradients there can be inherent difficulties training these quantum circuits. The alleviation of these difficulties is ongoing, with a good survey available \cite{fedorov2022vqe}, the references therein, as well as \cite{you2022convergence,PRXQuantum.2.020310,matsuo2023enhancing,PhysRevResearch.6.013254}.

\begin{figure}
\centering
\begin{quantikz}[column sep=5.5]
\lstick[3]{$\ket{\psi_{in}}$}&\gate{R_x(\theta_1)}&\gate{R_y(\theta_4)}&\gate{R_z(\theta_7)}&\ctrl{1}&\ctrl{2}&\qw&\qw&\rstick[3]{$\ket{\psi_{\boldsymbol{\theta}}}$,}\\
&\gate{R_x(\theta_2)}&\gate{R_y(\theta_5)}&\gate{R_z(\theta_8)}&\targ{}&\qw&\ctrl{1}&\qw&\\
&\gate{R_x(\theta_3)}&\gate{R_y(\theta_6)}&\gate{R_z(\theta_9)}&\qw&\targ{}&\targ{}&\qw&\\
\end{quantikz}
\caption{A single layer of the VQE ansatz that evolves an initial state, $\ket{\psi_{in}}$ to a parameterized state $\ket{\psi_{\boldsymbol{\theta}}} = \mathcal{U}(\boldsymbol{\theta})\ket{\psi_{in}}$. The output state is then used as a trial for the ground state and the expectation value of the Hamiltonian is computed. The parameters, $\boldsymbol{\theta}$, which are angles of the rotation gates, are varied to minimize this expectation value. Diagram created using \cite{kay2023tutorial}.}
\label{fig:VQE_circuit_1_layer}
\end{figure}

\subsection{Non-stabilizerness} \label{sec:non-stab}

The non-stabilizerness, also commonly called magic, is known to be the resource for quantum computation \cite{Chitambar_2019}. It is known, from the Gottesman-Knill theorem \cite{gottesman1998}, that Clifford circuits (that is, circuits comprised entirely of gates in the Clifford group) can be efficiently simulated on classical hardware. Therefore, in order to achieve a quantum advantage we need to go beyond the Clifford group, which means adding non-Clifford operations. Adding non-Clifford operations means increasing non-stabilizerness. The non-stabilizerness is measured by different methods, such as the robustness of magic \cite{veitch_mousavian_gottesman_emerson_2014, Howard_2017, hamaguchi2023handbook}, the min-relative entropy \cite{Liu2022} and stabilizer entropies \cite{Leone_2022,haug2023efficient}. 

For the purpose of this work, we use the 2-Renyi stabilizer entropy, $M_2$, since it is easily calculable for small systems and it has been shown to be a magic monotone  \cite{Haug2023stabilizerentropies,leone2024stabilizerentropiesmonotonesmagicstate}. The 2-Renyi stabilizer entropy for pure states is defined as
\begin{equation}
    M_2(\ket{\psi})=-\log \sum_{P\in \mathcal{P}_N}\frac{\bra{\psi}P\ket{\psi}^{4}}{2^N},
    \label{eq:SRE}
\end{equation}
where $\mathcal{P}_N$ is the set of all $N$-qubit Pauli strings and $\ket{\psi}$ can either be a parameterized representation of the quantum state, or the exact result found through exact diagonalization (ED). 

Recent studies have also explored the non-stablizerness in random circuits \cite{ahmadi2024quantifying,ahmadi2024mutual}, and many-body quantum systems such as the transverse-field Ising model \cite{PhysRevA.106.042426,10.21468/SciPostPhys.15.4.131,haug_mps_2023,PRXQuantum.4.040317}, the Potts model \cite{Tarabunga_2024,White_2021}, chaotic models \cite{passarelli2024chaosmagicdissipativequantum}, and the class of models known as generalized Rokhsar-Kivelson systems \cite{Tarabunga_2024}. It has also been shown that in special cases there can be a significant reduction of Equation \eqref{eq:SRE} into a form that scales only polynomially in the system size \cite{PhysRevA.110.022436}.

We are motivated by the observation that classical simulatability from the Clifford group point of view does not enter generic formulation of quantum or classical variational problems. We use Equation~\eqref{eq:SRE} to evaluate the non-stabilizerness of exact transverse-field Ising (TFI) ground states as well as its approximations determined using the methods described above, and compare the performance of these methods from the non-stabilizerness perspective across the phase diagram.

\section{Results} \label{sec:results}
\subsection{Quantum and Classical Model Accuracy} \label{sec:results-accuracy}
The simplest comparison between the three aforementioned methods is to compare them each to the results obtained from exact diagonalisation (ED).

\begin{figure}
    \centering
    \includegraphics{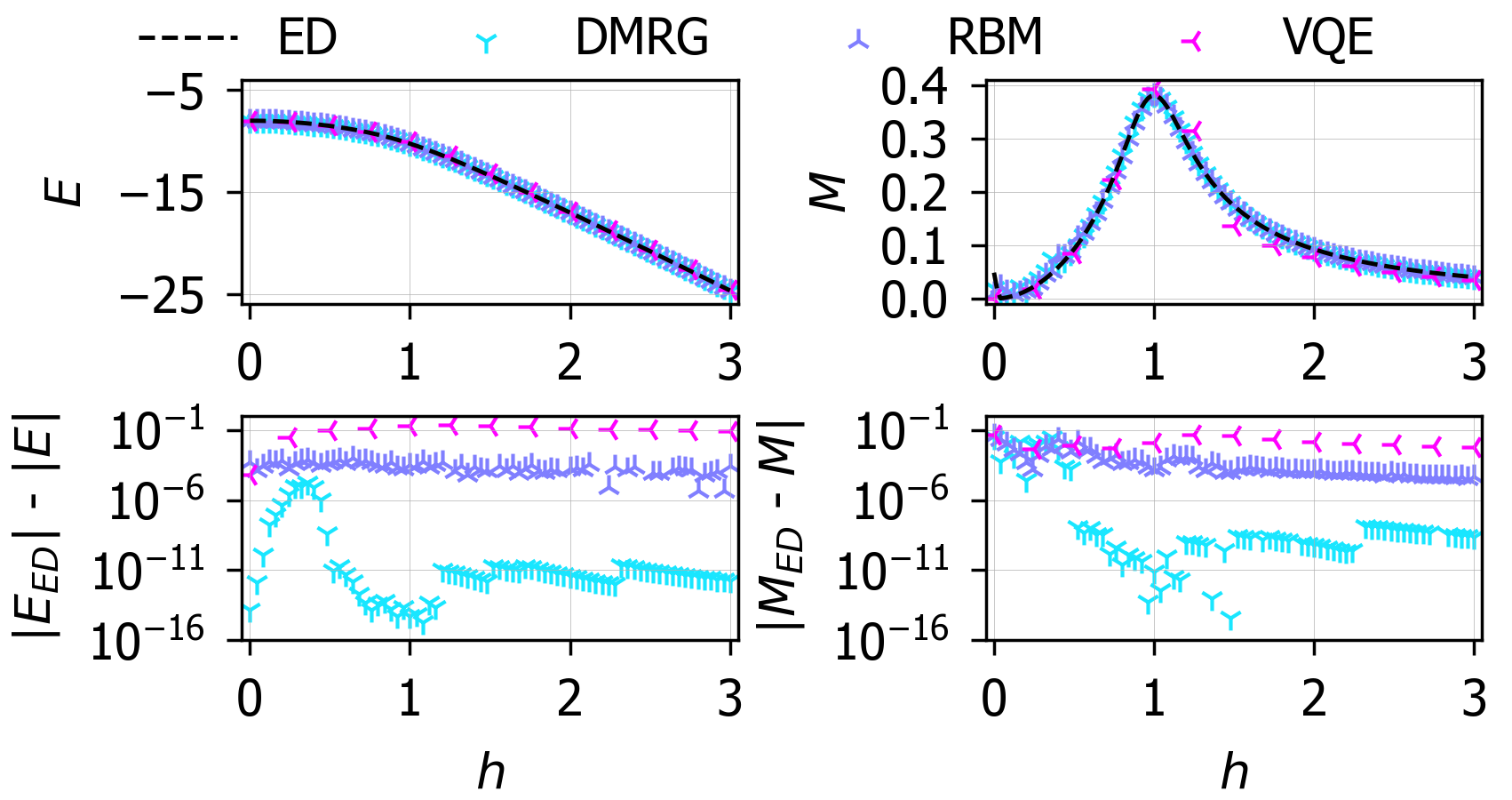}
    \caption{(top) The energy (left) and magic (right) of the ground state wavefunction for the 8 qubit transverse field Ising model at a range of transverse-field strengths, $h$. (bottom) The accuracy of the estimates of energy (left) and magic (right) from the three methods presented in Section \ref{sec:param_QS} relative to the result from exact diagonalization (ED). Each datum for the RBM and VQE is the mean of 10 repeats and the statistical errors are omitted for clarity; see Figure \ref{fig:statistical-errors} for error estimates.}
    \label{fig:8_qubit_best_of_each}
\end{figure}

Figure~\ref{fig:8_qubit_best_of_each} shows the energy and magic obtained from ED and all three variational approaches applied to an 8 qubit TFI system. Both the values themselves (top) and the accuracies with respect to the ED result (bottom) are shown. From the left-hand panels of this figure it is clear that there is a hierarchy in terms of energy accuracy: DMRG is more accurate than the RBM, which is in-turn more accurate than the VQE. From the right-hand panels, however, the magic accuracy does not quite follow this hierarchy; beyond $h=1$ the accuracies follow that of the energy, but approaching criticality from below there are signs that although the DMRG and RBM converged on states with more accurate energy than the VQE, the magic accuracies of all three are similar. It should be noted that in this region of the phase diagram the ground state of the TFIM is known to be degenerate (or near-degenerate, depending on finite system size effects and the specific value of $h$). The magic of the states within the degenerate eigenspace can differ greatly and thus the accuracy in energy does not constrain the state to have a similar magic to the ED result. One possible way to circumvent this would be to use ansätze that are aware of the symmetry that leads to this degeneracy. We will return to this in Section~\ref{sec:results-RBM-vs-symmetricRBM} as this highlights an important case of when energy accuracy alone is not sufficient for assessing the quality of the approximate ground state.

\begin{figure}
    \centering
    \includegraphics{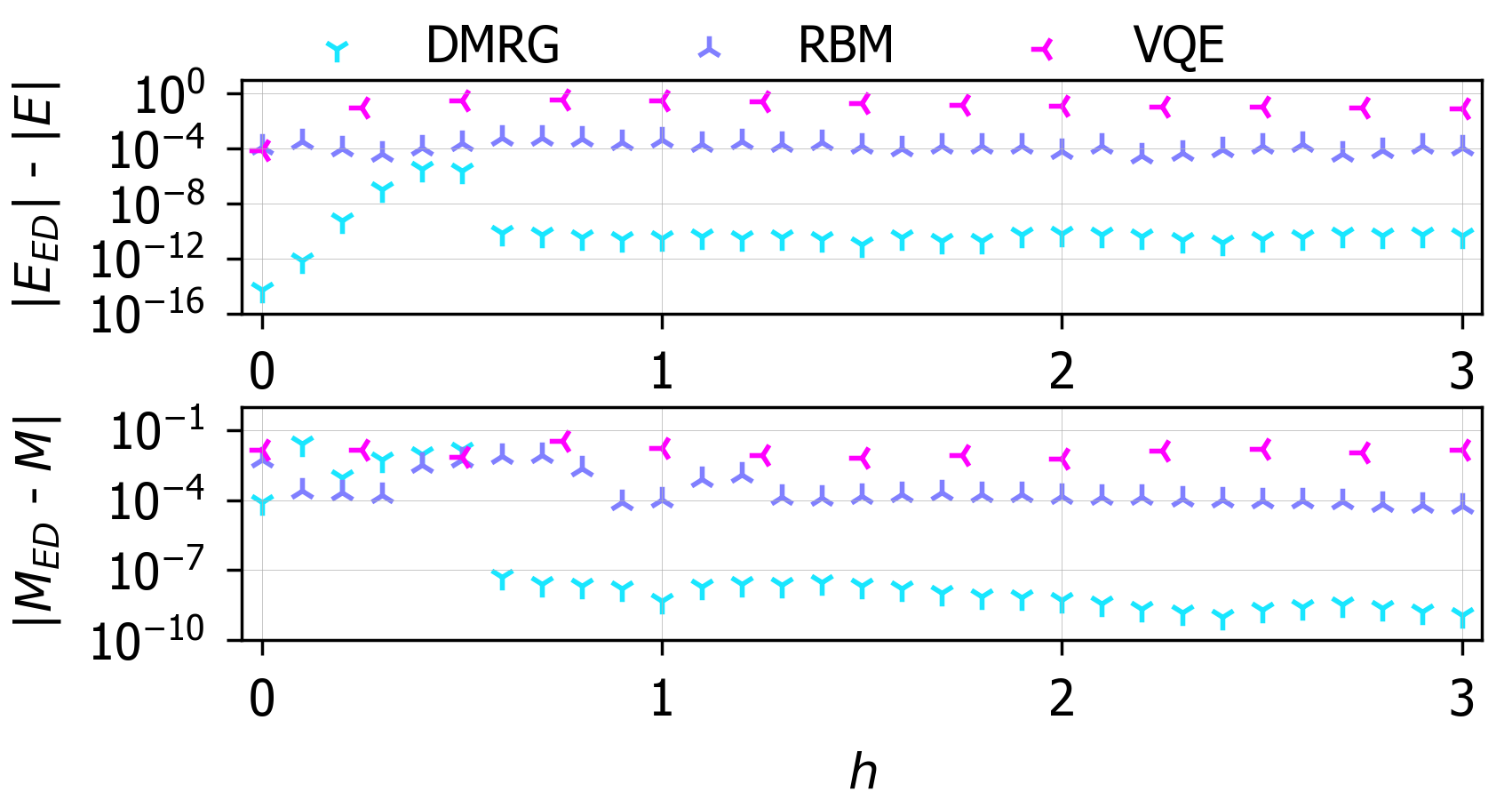}
    \caption{The accuracy of the estimates of energy (top) and magic (bottom) from the three methods presented in Section \ref{sec:param_QS} relative to the result from exact diagonalization (ED). These results are for a 12 qubit system, shown  at a range of transverse-field strengths, $h$. Each datum for the RBM and VQE is the mean of 10 repeats and the statistical errors are omitted for clarity; see Figure \ref{fig:statistical-errors} for error estimates.}
    \label{fig:12_qubit_best_of_each}
\end{figure}

Figure \ref{fig:12_qubit_best_of_each} shows the energy and magic accuracy of the same methods for a 12 qubit system. Due to the increased computational cost of simulating a larger system, the $h$ spacing is increased. Once again the accuracy in energy follows the ordering noted in Figure \ref{fig:8_qubit_best_of_each}. Also in a similar fashion to the 8 qubit simulation, the accuracy of the magic does not follow this trend over the entire phase diagram.

From Figures \ref{fig:8_qubit_best_of_each} and \ref{fig:12_qubit_best_of_each} we can conclude that energy accuracy and magic accuracy appear strongly correlated for most of the phase diagram, we can also say that, for these implementations, DMRG outperforms RBM which outperforms VQE when it comes to energy accuracy. In Appendix \ref{sec:VQE_appendix} we present the energy and magic accuracy computed for a 4 qubit system. In this smaller setting the VQE and RBM perform similarly. Despite the smaller system size, the same pattern in energy and magic accuracies can be seen: they appear correlated except for the low $h$ region. 

For clarity of the figures in this Section, the statistical fluctuations of the data were not shown. In Section~\ref{sec:results-fluctuations} will explore the statistical uncertainties to measure the robustness of these methods and further probe the correlations between the energy and magic of the ground state solutions found. 

\subsection{Non-stabilizerness and Symmetry} \label{sec:results-RBM-vs-symmetricRBM}

By and large, in the figures from Section \ref{sec:results-accuracy}, where we selected best performing hyperparameter-configuration for MPS, VQE and RBM, the accuracy of the energy appears to be directly linked to the accuracy of the magic. One may expect that by minimizing the energy towards the ground state then one ends up nearby in the Hilbert space and thus the magic of the solution is likely similar. However, the exception of the degenerate ground states at low $h$ was noted. This section will explore the interplay between the accuracy of energy and magic further, paying careful attention to the degeneracy of the solutions in the small $h$ region.

The Hamiltonian of the TFIM exhibits a global $\mathbb{Z}_2$ symmetry, meaning that the energy of two configurations that differ by flipping every single spin is the same. The ground states of this Hamiltonian do not have to respect this symmetry to still achieve the smallest possible energy. This leads to a range of different, but energetically equivalent, variational ground states. Importantly, the magic of these degenerate states can differ greatly. To give an indication of this, consider the $h=0$ case and any superposition of the all up and all down states (all of which have the same energy): an equal superposition has zero magic as it can be prepared using only Clifford operations acting on the all up state, however, for other unequal superpositions it is not guaranteed that one could prepare this state using only Clifford operations and thus it can have non-zero magic. To circumvent this, one can explicitly encode the global $\mathbb{Z}_2$ symmetry into the ansatz. We will explore this in the context of the RBM. 

For such a small, discrete symmetry group, one can make a symmetric (or $\mathbb{Z}_2$-invariant) RBM simply by feeding both the configuration $\textbf{s}$ and $-\textbf{s}$ through the RBM defined in Equation~\eqref{eq:RBM_amplitude} and taking the mean of each amplitude as the amplitude of the configuration $\textbf{s}$ (and thus also of $-\textbf{s}$). All other details of the training and RBM architecture are unchanged. 

To explore the intuition that energy accuracy alone is a good probe of whether or not the approximate ground state is similar to the true ground state, we compare it with two other figures of merit. Firstly, as in the other sections of this work, we compare with the magic accuracy, but then we also show the infidelity of the approximate ground state with the ground state found through ED. We define the infidelity of the normalized approximate state $\ket{\psi_{\boldsymbol{\theta}}}$ with the normalized ED result $\ket{\psi_{\text{ED}}}$ as
\begin{equation}
    I = 1 - |\langle \psi_{\boldsymbol{\theta}} | \psi_{\text{ED}} \rangle|^2.
\end{equation}

\begin{figure*}[ht]
    \centering
    \includegraphics{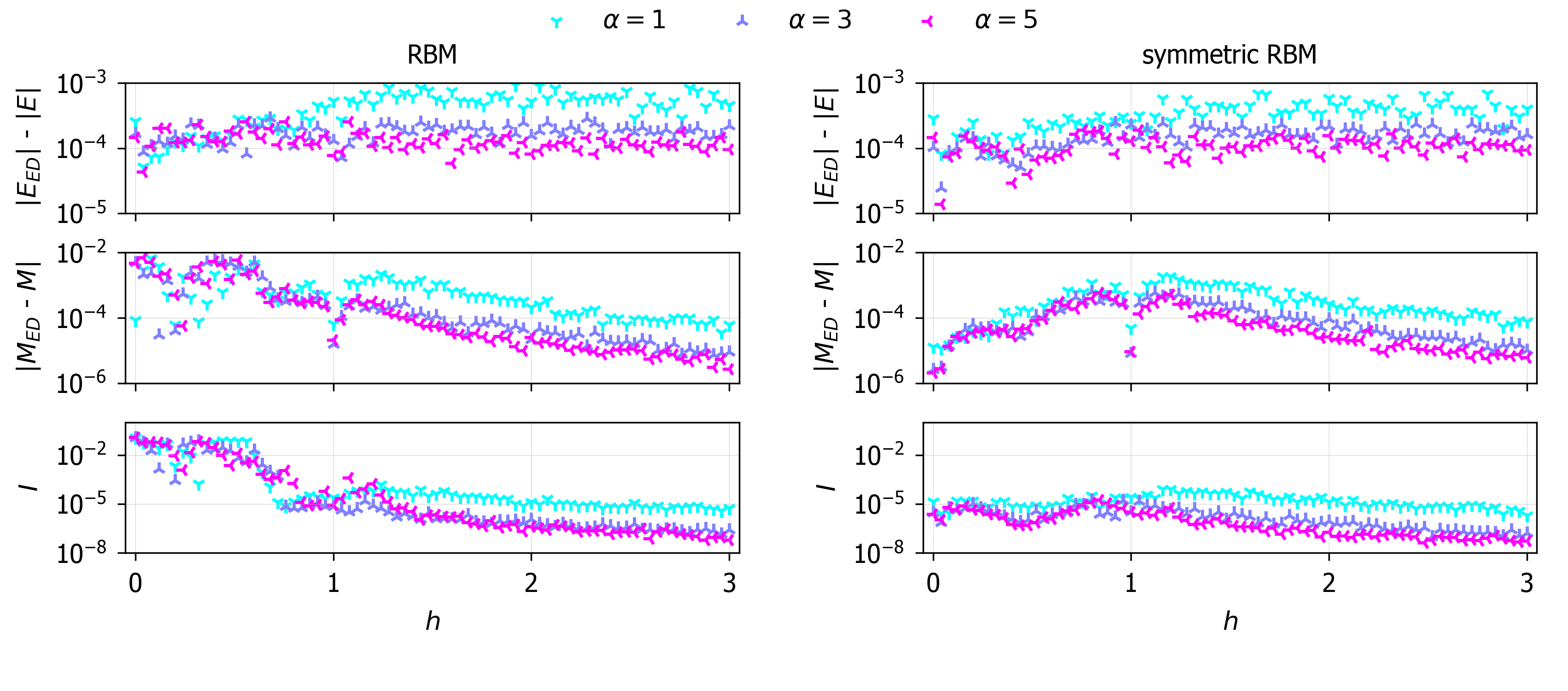}
    \caption{The energy accuracy (top), magic accuracy (middle), and infidelity (bottom) of the variational ground states compared to the ED results. The ansatz used for the NQS is either a conventional RBM defined in Equation~\eqref{eq:RBM_amplitude} (left) or the symmetric RBM as described in Section~\ref{sec:results-RBM-vs-symmetricRBM}  (right). These results are for an 8 qubit system, shown  at a range of transverse-field strengths, $h$, with each datum being the average of 10 repeats with the statistical errors omitted for clarity.}
    \label{fig:symmetric-vs-normalRBM}
\end{figure*}

Figure~\ref{fig:symmetric-vs-normalRBM} shows the energy accuracy, magic accuracy, and infidelity of the RBM and symmetric RBM ground states with respect to the ED results. The data are shown for three sizes of RBM, denoted by $\alpha = 1,3,$ and $5$, to explore if the increase in the number of parameters allows the network to learn this symmetry without the need for the explicit construction of the symmetric RBM. 

From Figure~\ref{fig:symmetric-vs-normalRBM} it is clear that the energy accuracy of the RBM and the symmetric RBM is similar. However, the effect on the magic accuracy and infidelity in the $0 \leq h \leq 1$ range is stark: the symmetric RBM outperforms the conventional RBM by up to five orders of magnitude. There does not seem to be much change to the rest of the $h$-range considered. Importantly, though, is that this five orders of magnitude improvement, indicating a significantly better ground state solution, is completely imperceptible when considering the energy accuracy alone.

The effect of increasing the size of the RBM, through increasing $\alpha$, is one of a mild increase in quality of all metrics, with the exception of the small $h$ region for the conventional RBM. Interestingly, this is the area that is most plagued by the effect of the degenerate ground states, and thus the extra parameters do not account for the symmetry. This means that just using a larger neural network was not enough to overcome the fact that energy accuracy alone is insufficient for the ansatz to fully approximate the non-stabilizerness of the ground state.

\subsection{Fluctuations in Quantum and Classical Solutions} \label{sec:results-fluctuations}

One theme that this work aims to shed some light onto is the representativeness of a ground state found through energy minimization alone. Magic is used as a second axis to ascertain how well the approximate state represents the true ground state. In this section we will explore how similar in magic repeated energy minimizations are, probing the landscape around, what one hopes to be, the energy minimum containing the true ground state.

\begin{figure}
    \centering
    \includegraphics{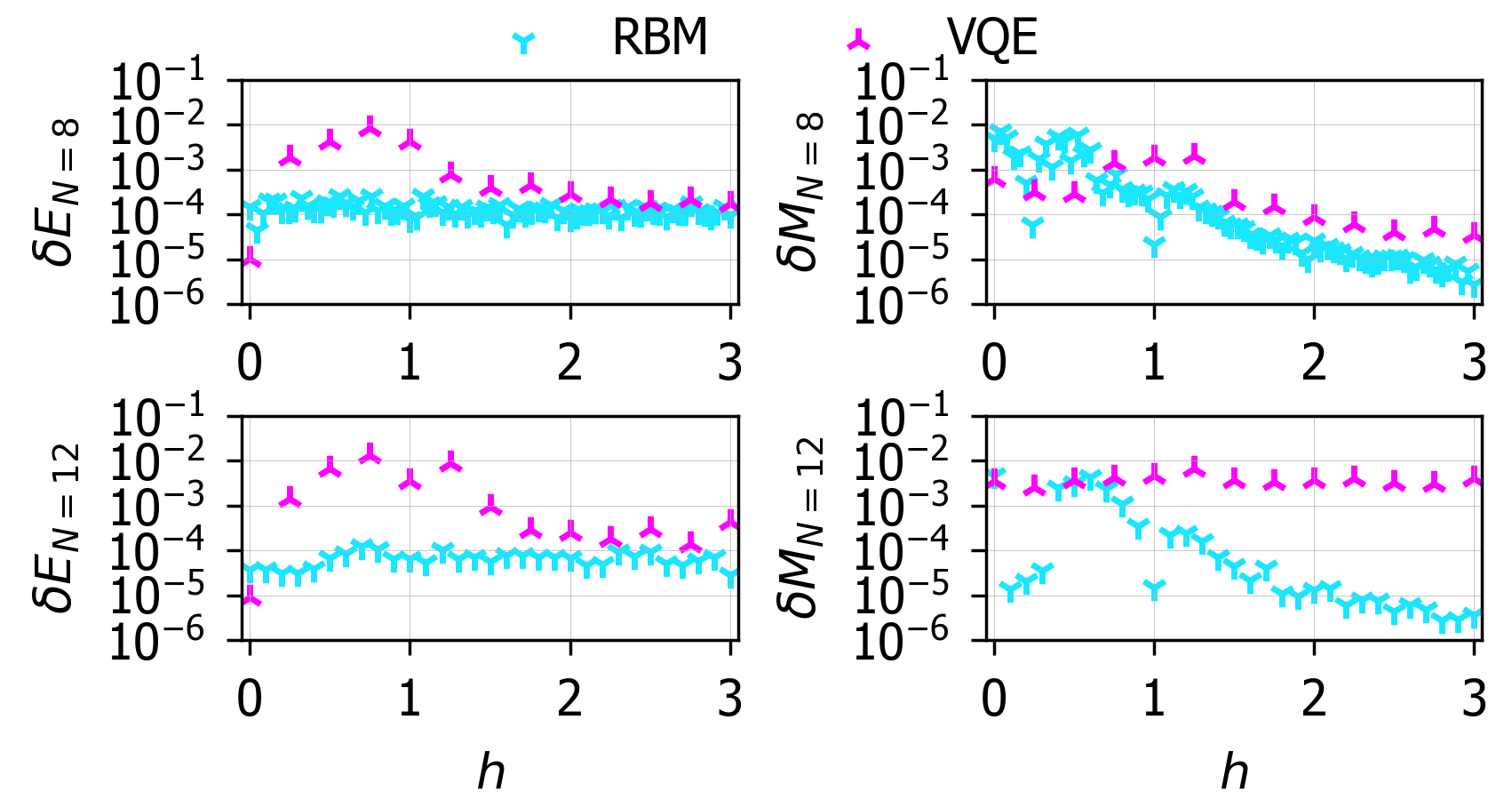}
    \caption{The statistical errors of the energy (left) and magic (right) against the transverse-field strength, $h$, across the 10 repeats of the ground state optimization procedure using an RBM and VQE. Data are shown for 8 (top) and 12 (bottom) qubits, with the corresponding mean values reported in Figures \ref{fig:8_qubit_best_of_each} and \ref{fig:12_qubit_best_of_each} respectively.}
    \label{fig:statistical-errors}
\end{figure}

Figure \ref{fig:statistical-errors} shows the statistical error from 10 repeated ground state searches, in energy (left) and magic (right), for 8 qubit (top) and 12 qubit (bottom) simulations. The comparison is shown only for the VQE and RBM methods. From this figure it can be seen that the fluctuations in energy are almost always smaller for the RBM compared to the VQE. The exception to this is the $h=0$ point; the ground state of the Hamiltonian here is, in fact, the initial state fed into the VQE and thus this is perhaps not surprising. The fact that the VQE can, in very high and low field, reach almost the level of consistency of the RBM yet it cannot minimize the energy as well suggests that perhaps this is a limitation of the expressivity of the ansatz. 

In Section~\ref{sec:results-RBM-vs-symmetricRBM} we already noted the difficulties with an RBM that is not aware of the symmetries of the system. This is reflected here in the fluctuations of the magic; interestingly this is less pronounced in the 12 qubit simulation. Those data notwithstanding, the fact that the fluctuations of the magic are very small does suggest that each approximate ground state, when near to the minimum in energy, is also in a region with similar magic.

\section{Discussion and Conclusions}
\label{sec:conclusions}

In this work we set up the framework to reconcile classical simulatability notions of quantum resource theory with the formalism of classical variational techniques. We used non-stabilizerness, a quantity that measure how far is a given state from being classically simulatable in a sense of Gottesmann-Knill theorem, as a figure of merit for quality of classical variational approximation of a quantum state. Specifically, we assessed non-stablizerness expressivity for three qualitatively different types of variational ansätze: tensor networks, neural networks and variational quantum circuits.

We found, that when comparing the best model (in terms of energy performance) for each method with exact diagonalization, on average, there is a general trend for better energy to correspond to better approximation in non-stabilizerness. This is an encouraging observation that suggests that the state reconstructed as a result of classical variational procedure has a complexity structure that represents the exact quantum solution reasonably well. At the same time, we immediately noted that this correspondence is far from straightforward and universal. For example, an RBM constructed to share the symmetry of the Hamiltonian led to ground states with up to four orders of magnitude higher accuracy in non-stabilizerness than a conventional RBM, with no change in the energy accuracy.

We hope this work will be a stepping stone towards starting a closer dialogue between quantum computing and classical variational perspectives on solving the quantum many-body problem, and more specifically, on how the quality of the results is assessed. We set up a system of benchmarks for small system sizes ($N=8$,$12$) and noted a number of similarities and differences between accuracies in non-stabilizerness and in energy. The possible next steps include obtaining large scale benchmarks for each of the methods we tested here. The current limitation of system size is the calculation of the non-stabilizerness, however, the first impressive step towards this calculation at scale has already been taken in Ref.~\cite{tarabunga2024nonstabilizerness} for matrix product state representation of a wavefunction. It would be interesting to see how the smaller system observations translate into a large scale benchmark. Further ahead, one could think of how to embed non-stabilizerness optimization iterations into energy-based quantum and classical variational models.

\section{Data and Code Availability}
A GitLab repository containing this project is available at~\cite{gitlab}. All the data and code to analyze them is available at~\cite{zenodo}. 

\section{Author Contributions}
EG designed the project with input from TS and AA. TS and AA jointly performed neural network and ED simulations. TS and BC performed the VQE simulations. TS performed the tensor network simulations. AA provided the non-stabilizerness calculation. TS, AA, BC, and EG analyzed and interpreted the data. TS, EG, and AA co-wrote the manuscript. EG supervised the project.

\section{Acknowledgments}

We are thankful for instructive discussions on DMRG with B. M. La Rivi\`ere, J. Soto Garcia, and P. S. Tarabunga. We also thank A. Valenti, J. Nys, and S. Khaleefah for fruitful conversations regarding NQS. This work is part of the project Engineered Topological Quantum Networks (Project No.VI.Veni.212.278) of the research program NWO Talent Programme Veni Science domain 2021 which is financed by the Dutch Research Council (NWO). This publication is also part of the project Optimal Digital-Analog Quantum Circuits with file number NGF.1582.22.026 of the research programme NGF-Quantum Delta NL 2022 which is (partly) financed by the Dutch Research Council (NWO) and the Dutch National Growth Fund initiative Quantum Delta NL.

\appendix
\section{NQS architecture} \label{sec:RBM_appendix}

There are a vast number of hyperparameters and architectural configurations that one can tweak to alter the training of the NQS. To arrive at the specific RBM presented in Figures~\ref{fig:8_qubit_best_of_each} and \ref{fig:12_qubit_best_of_each} we explored: the size of the ansatz, using the entire Hilbert space for the computation of expectation values, and the inclusion of the stochastic reconfiguration (SR) preconditioner as part of the parameter optimization routine. Other avenues of exploration could include introducing a representation learning step to map the configurations from projective measurements to continuous vectors \cite{Viteritti:2022fji} or other neural network architectures \cite{carleo_RBM_2017,PhysRevB.100.125124, Hibat-Allah_Carrasquilla_2020, Viteritti:2022fji, Luo:2020stn, roth2021group}, but these lie outside the scope of this work. This appendix will also cover the stopping criterion used to estimate when the ground state is approximately reached. 

The effects of a larger RBM were shown in Figure~\ref{fig:symmetric-vs-normalRBM}, from this it can be seen that there is a mild improvement in all metrics with increased model size. The drawback, however, is in the time taken to train the larger model: more parameters leads to increased training time. Given that the improvement is larger from $\alpha=1$ to $\alpha=3$ than it is between $\alpha=3$ and $\alpha=5$, alongside the increased training time, we chose to stop at $\alpha=5$. 

In Section~\ref{sec:nqs} we mentioned the approximations made as part of the VMC procedure, notably, as part of the computation of the expectation value, the reduction of the sum over the entire Hilbert space (Equation~\eqref{eq:expectation}) to the Monte Carlo estimate sampled from Equation~\eqref{eq:MC_sampling_P}. We tested the energy and magic accuracy for an 8 qubit system using both the full expectation value and the Monte Carlo estimate to train the RBMs. We found that, across the entire range of $h$, the fluctuations between repeated optimizations of the exact expectation solutions were lower than the Monte Carlo sampled counterparts, and the energy accuracies were smoother as a function of $h$ for the exact expectation method, too. Both effects were, however, mild, and as the exact expectation is much more computationally expensive (it scales exponentially with the system size), we chose to remain with the Monte Carlo sampling.

Finally, for the configuration of the RBM, we tested the inclusion of the SR preconditioner. Introduced to VMC in \cite{PhysRevLett.80.4558,PhysRevB.71.241103}, SR alters the gradient obtained from an optimizer in a way that accounts for the curvature of the optimization landscape that is caused only by the ansatz. This can be seen as an application of what is known as the natural gradient in the wider machine learning community \cite{10.1162/089976698300017746,Stokes2020quantumnatural}. Put simply, the inclusion of SR lead to better energy and magic accuracies of the ground states found.

During the optimization of the RBM parameters one must establish a point at which to stop. For this work we considered the training to be done when the relative change in energy did not improve by more than $10^{-7}$ for 500 consecutive epochs. This can be easily implemented using the \texttt{callbacks.EarlyStopping} function from \texttt{NetKet}.

We used the ADAM \cite{Kingma:2014vow} optimizer and 1,000 (5,000) samples for the MC estimates of expectation values for the 8 (12) qubit systems for all of the RBM results presented in this work.

\section{VQE architecture} \label{sec:VQE_appendix}

Whilst, in theory, VQE is a Hamiltonian-agnostic and flexible algorithm, there are optimizations that can be made in its implementation. In a similar fashion to Appendix \ref{sec:RBM_appendix}, this section will explore the choices made when exploring the implementation of the VQE: the number of layers in the ansatz, the method for computing the expectation value of the Hamiltonian, and the optimizer. It will also elucidate the stopping criteria used.

Figure~\ref{fig:VQE_circuit_1_layer} shows a single layer of the VQE ansatz used in this work. We explored using up to four layers of this as the ansatz and found that the energy and magic accuracies improved with more layers, however, interestingly the performance of one and two layers were similar, as were three and four layers. The training time increases with each added layer and therefore we chose not to go beyond four layers, using four layers as the final ansatz. 

Given that this VQE implementation was performed using a classical simulation of a quantum circuit, we could compute the expectation value of the Hamiltonian two ways: either using projective measurements of the final state or by exact computation (exploiting the full wavefunction of the final state that is only easily accessible in classical simulations and not in real quantum experiments). We tested using both methods to compute the expectation value of the Hamiltonian during training and found no significant difference between the two sets of ground states found, neither in energy nor magic. However, the time taken for the ground state search when using measured expectations was up to three orders of magnitude longer than when using exact expectations (up from $10^2$ to $10^5$ seconds). This could be caused by several things, the two that seem most pertinent to mention are related to the gradient of the expectation of the Hamiltonian with respect to the circuit parameters. Firstly, as the projective measurement step of a quantum circuit is not differentiable, the gradient was estimated using the parameter-shift rule \cite{PhysRevA.98.032309,PhysRevA.99.032331} which requires multiple evaluations of the circuit; this is in contrast to the exact case which only required a single execution to compute the value and gradient of the expectation value. Secondly, as an estimation of the expectation value can only be as accurate, or less accurate, than the exact value, the estimations of the gradient of the expectation value with respect to the circuit angles can only be equal to, or less accurate, than the exact case. Given this, then, it is natural to assume that with likely many sub-optimal estimations of the gradient, the optimization procedure would take longer and require more evaluations of the quantum circuit. Both of these factors will contribute to the increased runtime of the VQE procedure. Given the lack of clear improvement by either approach, and as this paper does not extend to system sizes beyond the range of what is classically simulatable, we chose to use the exact expectation value for the VQE procedure for significantly more efficient use of computational resources. 

To assess the point at which we stop training the quantum circuit we implemented a two-step stopping criteria. Firstly, to consider a single instance of a ground state to have been found we require that three consecutive epochs give the same energy to within $10^{-6}$, but then we also require that if the whole VQE is reinitialised then the next instance must not differ from the previous by more than a relative change of $10^{-4}$ in the energy. The training of the quantum circuit was less stable and led to larger fluctuations in energy than for the RBM, and thus the stopping criterion for the single instance is less strict, but the problem of local minima in the search space was more common here and thus the second part of the stopping criteria was added.

\begin{figure}
    \centering
    \includegraphics{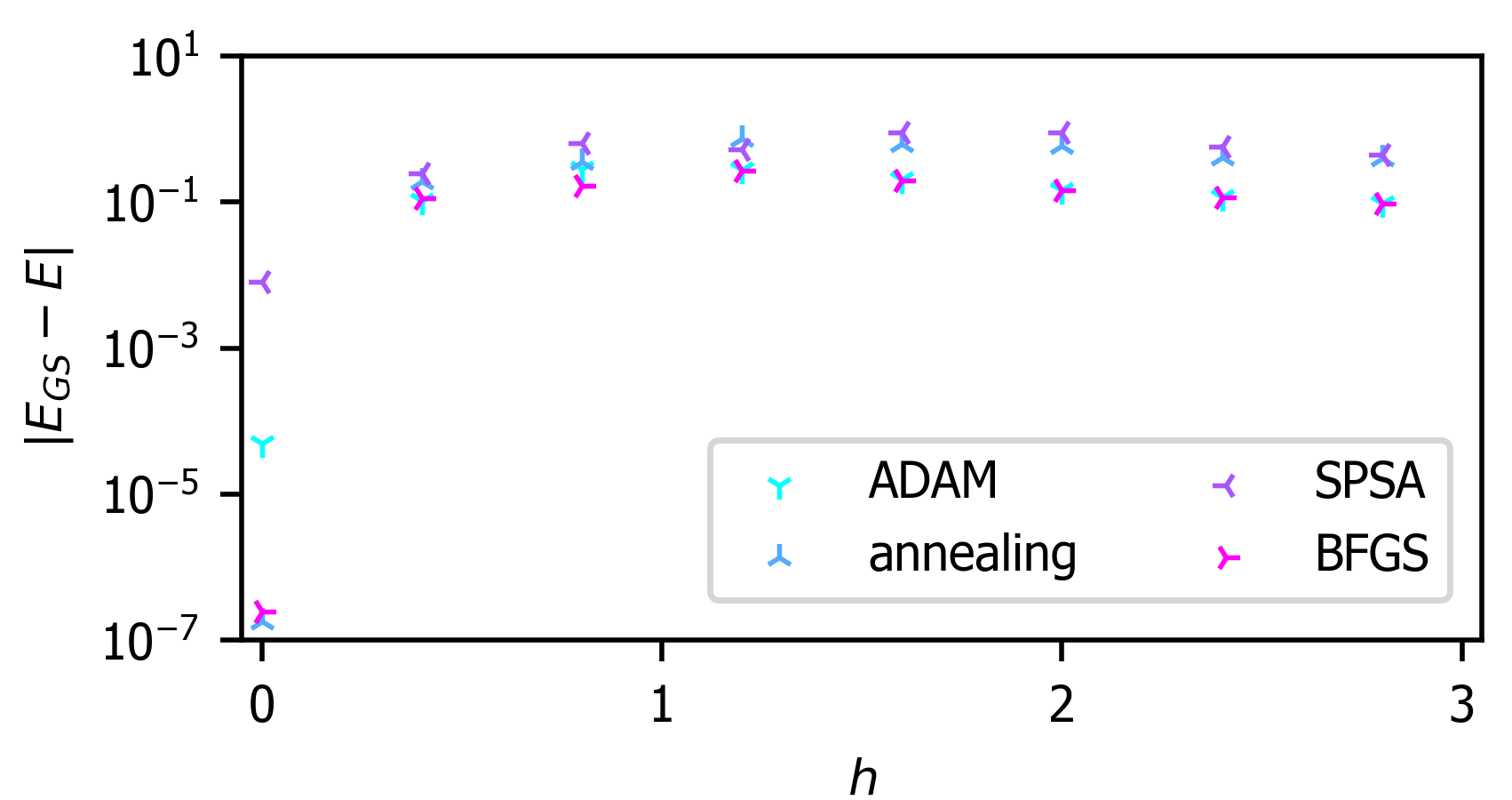}
    \caption{Energy accuracy against transverse-field strength for a three-layer VQE ansatz, shown for a range of optimizers used in the energy minimization procedure. All data are for an 8 qubit simulation.}
    \label{fig:VQE_optimizers}
\end{figure}

\begin{figure}
    \centering
    \includegraphics{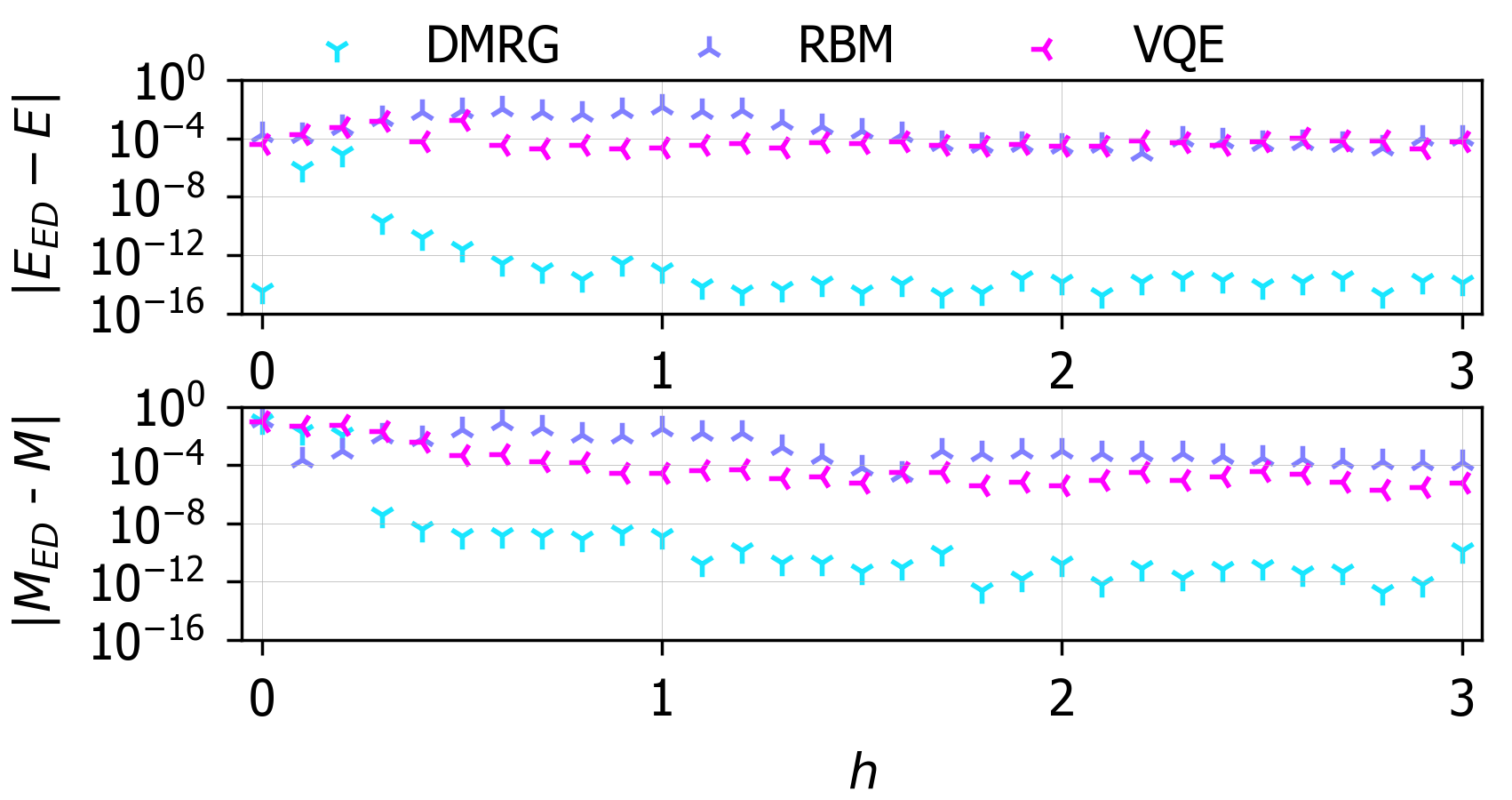}
    \caption{The energy accuracy (top) and magic accuracy (bottom) against transverse-field strength for a 4 qubit simulation. Each datum is the average over 10 minimizations.}
    \label{fig:4_qubit_best_of_each}
\end{figure}

There is a large community interest in alleviating the problem of barren plateaus in quantum machine learning applications. Specifically to VQEs, the authors in \cite{PhysRevResearch.2.043246} explored the effect of using three different optimizers: Broyden-Fletcher-Goldfarb-Shanno (BFGS) \cite{10.1093/imamat/6.1.76,10.1093/comjnl/13.3.317,goldfarb1970family,shanno1970conditioning}, ADAM \cite{Kingma:2014vow}, and the natural gradient \cite{Stokes2020quantumnatural,10.1162/089976698300017746} to avoid the problem of local minima in the variational landscape. Here we consider the energy accuracy of an 8 qubit energy minimization of a VQE for four different optimizers: ADAM, BFGS, SPSA \cite{spall1992multivariate}, and simulated annealing \cite{xiang1997generalized}. This is a mix of gradient- and non-gradient-based optimizers: BFGS and ADAM were chosen to cover those of interest for avoiding local minima, SPSA for its utility with applications on quantum hardware \cite{kandala2017hardware}, and annealing as a second non-gradient based approach as this avoids the use of the parameter-shift rule. The ansatz chosen for these optimizations is the three-layer version of the ansatz depicted in Figure \ref{fig:VQE_circuit_1_layer}. The number of layers was reduced to three because the annealing optimizer consistently failed to converge for a four-layer ansatz. Figure \ref{fig:VQE_optimizers} shows the energy accuracies achieved from 10 repeated energy minimizations of the TFI Hamiltonian for each of the chosen optimizers. From this we can see that both BGFS and ADAM perform the best but there is a very quick degradation of performance with increased transverse field strength, $h$. As concluded by the authors of Ref.~\cite{PhysRevResearch.2.043246}, both ADAM and BFGS struggle to scale to systems with a large number of parameters; for the ansatz used here there are 72 free parameters, much larger than the upper limit of 42 that they propose. BFGS especially struggles to scale to larger systems. For this reason,  all other VQE data shown in this work uses ADAM as the optimizer. As previously mentioned, reducing the number of layers in the ansatz led to worse energy accuracies, so we will now explore the other way to reduce the number of parameters: a smaller system size.

To explore a regime where the VQE optimization maintains a flexible and expressive ansatz, but with fewer variational parameters, we will consider a 4 qubit TFI system. Exactly as done in Section \ref{sec:results-accuracy}, Figure \ref{fig:4_qubit_best_of_each} shows the accuracies in energy and magic of the VQE, RBM, and DMRG algorithms. Unlike in Figures \ref{fig:8_qubit_best_of_each} and \ref{fig:12_qubit_best_of_each}, however, the energy accuracy of the VQE is better than that of the RBM. This shows that the VQE is able to achieve accurate ground state energies, but Figures \ref{fig:8_qubit_best_of_each} and \ref{fig:12_qubit_best_of_each} show that this does not easily scale to larger systems with our current implementation. It should also be noted that in Figure \ref{fig:4_qubit_best_of_each} we can still see that the accuracy in energy does not always correlate with accuracy in magic.

\bibliographystyle{SciPost_bibstyle}
\bibliography{mybibliography}

\end{document}